\documentclass[a4paper]{jpconf}
\usepackage{graphicx,graphics,amsfonts,amsmath,amssymb,bm,hyperref,psfrag}

\newcommand{\be}{\begin{equation}}
\newcommand{\ee}{\end{equation}}

\begin{document}

\title{Gravitational-wave memory revisited: memory from the merger and recoil of binary black holes}
\author{Marc Favata}
\address{Kavli Institute for Theoretical Physics, University of California, Santa Barbara, CA 93106-4030}
\ead{favata@kitp.ucsb.edu}
\begin{abstract}
Gravitational-wave memory refers to the permanent displacement of the test masses in an idealized (freely-falling) gravitational-wave interferometer. Inspiraling binaries produce a particularly interesting form of memory---the Christodoulou memory. Although it originates from nonlinear interactions at 2.5 post-Newtonian order, the Christodoulou memory affects the gravitational-wave amplitude at leading (Newtonian) order. Previous calculations have computed this non-oscillatory amplitude correction during the inspiral phase of binary coalescence. Using an ``effective-one-body'' description calibrated with the results of numerical relativity simulations, the evolution of the memory during the inspiral, merger, and ringdown phases, as well as the memory's final saturation value, are calculated. Using this model for the memory, the prospects for its detection are examined, particularly for supermassive black hole binary coalescences that LISA will detect with high signal-to-noise ratios.  Coalescing binary black holes also experience center-of-mass recoil due to the anisotropic emission of gravitational radiation. These recoils can manifest themselves in the gravitational-wave signal in the form of a ``linear'' memory and a Doppler shift of the quasi-normal-mode frequencies. The prospects for observing these effects are also discussed.
\end{abstract}
\section{\label{sec:intro}Introduction and motivation}
In the standard picture of the gravitational-wave (GW) signal from a coalescing binary, the oscillating amplitude starts small at early times, grows to some peak amplitude, and than decays back to its zero-value at late times. For example, the dominant ($l=m=2$) mode of the GW polarizations from coalescing binary black holes (BBHs) follows this standard picture. However, some sources exhibit a gravitational-wave \emph{memory} in which the values of the GW polarization amplitudes differ at late and early times:
\be
\label{eq:memdef}
\Delta h_{+,\times}^{\rm mem} = \lim_{t\rightarrow +\infty} h_{+,\times}(t) - \lim_{t\rightarrow -\infty} h_{+,\times}(t),
\ee
where $t$ is the observer's time.
In an ideal GW detector (eg., a ring of freely-falling test masses), the memory causes a permanent displacement that persists after the GW has passed.

Gravitational-wave memory comes in two types: The \emph{linear memory} has been known since the 1970's (\cite{braginskii-thorne} and references therein) and arises from sources that produce a net change in the time-derivatives of their \emph{source-multipole-moments}.
A simple example of a source with linear memory is a binary on a hyperbolic orbit (gravitational two-body scattering) \cite{turner-unbound}. To see how memory arises in this system, consider the leading-order transverse-traceless (TT) GW field, $h_{jk}^{\rm TT} \propto \ddot{\mathcal I}_{jk}^{\rm TT}/R$, where $R$ is the distance to the observer, ${\mathcal I}_{jk}=\eta M [x_i x_j]^{\rm STF}$ is the source quadrupole moment, $M$ is the total mass, $\eta$ is the reduced-mass ratio, $x_j$ is the relative binary separation vector, and STF means to take the symmetric-trace-free part. Using the Newtonian equations of motion, one can easily see that at large separations (before and after scattering) $\ddot{x}_j \rightarrow 0$, but $\dot{x}_j$ is a constant for unbound systems and has a different direction at late and early times. This leads to a memory $\Delta h_{jk}^{\rm TT} \propto  \eta M \Delta [\dot{x}_j \dot{x}_k]^{\rm TT}/R$.

Other systems with linear memory are those whose components change from being bound to unbound (or vice versa). This includes binaries whose members are captured, disrupted, or undergo mass loss. Gravitational-waves with linear memory have been studied in the context of supernova explosions
(see \cite{ott-corecollapsereview} for a review and references), mass loss due to neutrino emission \cite{epstein-neutrinomemory}, and gamma-ray-burst jets \cite{sago-GRBmemory}.
A general formula for the linear memory is given by \cite{kipmemory,braginskii-thorne}:
\be
\label{eq:hijlinmem}
\Delta h_{jk}^{\rm TT} = \Delta \sum_{A=1}^N \frac{4 M_A}{R\sqrt{1-v_A^2}} \left[ \frac{v_A^j v_A^k}{1-{\bm v}_A \cdot {\bm N}} \right]^{\rm TT} ,
\ee
where $A$ is an index that labels the $N$ masses $M_A$ with velocities ${\bm v}_A$ that are unbound in their initial or final states (or both), and ${\bm N}$ is a unit vector that points from the source to the observer.

The \emph{nonlinear memory} was discovered independently by Payne \cite{payne-zfl},  Blanchet and Damour \cite{blanchet-damour-hereditary}, and Christodoulou \cite{christodoulou-mem}. It is often referred to as the ``Christodoulou memory.''  The nonlinear memory arises from a change in the \emph{radiative multipole moments} that is sourced by the energy-flux of the radiated GWs. One can heuristically understand the origin of the nonlinear memory as follows: Consider the relaxed Einstein field equations (EFE) in harmonic gauge: $\Box \bar{h}^{\alpha \beta} = -16 \pi \tau^{\alpha \beta}$, where $\bar{h}^{\alpha \beta}$ is the gravitational field tensor and $\tau^{\alpha \beta}$ depends on the matter stress-energy tensor, the Landau-Lifshitz pseudotensor $t_{\rm LL}^{\alpha \beta}$, and other terms quadratic in $\bar{h}^{\alpha \beta}$ \cite{kiprmp}. Of the many nonlinear terms in $t_{\rm LL}^{\alpha \beta}$, there is a particular piece that is proportional to the stress-energy tensor for GWs:
$T^{\rm gw}_{jk} = \frac{1}{R^2} \frac{dE^{\rm gw}}{dt d\Omega} n_j n_k$,
where $\frac{dE^{\rm gw}}{dt d\Omega}$ is the GW energy flux and $n_j$ is a unit radial vector.  When applying the standard Green's function to the right-hand-side of the relaxed EFE, this piece yields the following correction term to the GW field \cite{wiseman-will-memory}:
\be
\label{eq:hTTnonlinmem}
\delta h^{\rm TT}_{jk} = \frac{4}{R} \int_{-\infty}^{T_R} dt'\, \left[ \int \frac{dE^{\rm gw}}{dt' d\Omega'} \frac{n'_j n'_k}{(1-{\bm n}' \cdot {\bm N})} d\Omega' \right]^{\rm TT},
\ee
where $T_R$ is the retarded time. This shows that part of the distant GW field is sourced by the loss of GW energy.
Thorne \cite{kipmemory} has shown that the nonlinear memory [Eq.~\eqref{eq:hTTnonlinmem}] can be described in terms of the linear memory [Eq.~\eqref{eq:hijlinmem}] if the unbound objects in the system are taken to be the individual radiated gravitons with energies $E_A=M_A/(1-v_A^2)^{1/2}$ and velocities $v^j_A = c \, n_{A}'^j$.

The Christodoulou memory is a particularly interesting manifestation of the nonlinearity of general relativity. Although it arises from multipolar interactions beginning at the 2.5-post-Newtonian (PN) order \cite{arun25PNamp}, the nonlinear memory affects the GW amplitude at leading (Newtonian) order. Like GW tails the memory depends on the entire past-history of the source; but unlike most other nonlinear PN effects, the memory is non-oscillatory and causes a slowly-growing shift in the $+$ polarization\footnote{For circular orbits $h_{\times}$ is unaffected by the Christodoulou memory. However, during the inspiral there is a DC contribution to $h_{\times}$ at 2.5PN order from nonlinear corrections to the radiative current octupole moment \cite{arun25PNamp}.}.

Here we focus on the memory from merging BBHs. While the oscillatory pieces of the GW polarizations during the inspiral are known to 3PN order \cite{blanchet3pnwaveform}, the Christodoulou memory has, until recently, only been calculated to leading-(Newtonian)-order\footnote{Blanchet et.~al \cite{blanchet3pnwaveform} have shown that the 0.5PN correction to the memory is exactly zero for quasi-circular orbits. The 3PN corrections to the memory pieces of the inspiral waveform are reported in Ref.~\cite{favata-pnmemory}.} \cite{wiseman-will-memory,arun25PNamp}. However a proper determination of the memory's detectability requires not only knowing how it slowly accumulates during the inspiral, but also how the memory rapidly grows and saturates to its final value during the merger and ringdown. Earlier estimates of the memory's detectability have either made crude order-of-magnitude estimates \cite{kipmemory} or considered only the memory during the inspiral \cite{kennefick-memory}.

In principle numerical relativity (NR) simulations can compute the memory at all stages of the coalescence. In practice extracting the memory from these simulations faces several difficulties: NR simulations can best resolve the dominant $l=m=2$ spin-weighted spherical-harmonic mode of the $\Psi_4$ curvature scalar; but for circular orbits the memory is present only in the $m=0$ modes, which are smaller by 5PN orders (or about four orders-of-magnitude) during the late inspiral.  The memory is also sensitive to the two integration constants that must be determined when computing the metric perturbations from the curvature perturbations. For simulations that compute the metric perturbations directly, the $m=0$ modes enter at the same PN order as the $l=m=2$ mode, but are still numerically smaller by nearly three orders of magnitude. The sensitive dependence of the memory to the binary's past history means that large errors can result unless the simulations start with large initial binary separations. See Ref.~\cite{favata-pnmemory} for further discussion.
In the absence of results from NR, the purpose of this work is to estimate the evolution and saturation of the memory during the merger and ringdown phases.
\section{\label{sec:memcalc}Calculating the memory from binary black hole coalescence}
The leading-order PN multipole moment expansion of the oscillatory and memory pieces of the GW polarizations can be expressed as \cite{favata-memory-saturation}:
\be
\label{eq:h0}
h_{+}^{(0)} - i h_{\times}^{(0)} \approx \frac{1}{8R}\sqrt{\frac{5}{2\pi}} \left[ (1+\cos\Theta)^2 e^{2i\Phi} I^{(2)}_{22} + (1-\cos\Theta)^2 e^{-2i\Phi} I_{2 -2}^{(2)} \right] , \;\;\; \text{and}
\ee
\be
\label{eq:hmem}
h_+^{\rm mem} \approx \frac{\eta M {h}^{\rm mem}}{384 \pi R} \sin^2\Theta (17 + \cos^2\Theta), \;\; {h}^{\rm mem} \equiv \frac{1}{\eta M} \int_{-\infty}^{T_R} |I^{(3)}_{22}(\tau)|^2 d\tau \approx \frac{16\pi}{\eta} \left(\frac{\Delta E_{\rm rad}}{M}\right).
\ee
Here $(\Theta,\Phi)$ are the direction to the observer and $I^{(n)}_{2 \pm2}$ is the $n^{\rm th}$-time-derivative of the spherical harmonic coefficient of the source mass-quadrupole moment\footnote{These ``scalar'' moments are related to the more familiar STF-source-quadrupole moment ${\mathcal I}_{ij}$ via $I_{2m} = (16 \pi \sqrt{3}/15) {\mathcal I}_{ij} {\mathcal Y}_{ij}^{2m\, \ast}$, where the ${\mathcal Y}_{ij}^{2m}$ are related to the ordinary spherical harmonics by $Y^{2m} = {\mathcal Y}_{ij}^{2m} n^i n^j$ \cite{kiprmp}.}.

To model the evolution of the source-quadrupole moment we restrict ourselves to non-spinning, circularized BBHs and follow the ``effective-one-body'' (EOB) approach (see \cite{EOB-damour-lecnotes} and references therein), calibrated to the results of NR simulations. Before using the full EOB formalism, it is instructive to first consider a bare-bones version of EOB called the \emph{minimal waveform model} (MWM). The MWM is a simple, analytic model for the inspiral, merger, and ringdown that qualitatively captures some of the important physics while minimizing complexity. It consists of modeling the multipole moments during the inspiral by their leading-order PN expressions and then matching to a sum of quasi-normal modes (QNMs). During the inspiral the source quadrupole-moment derivatives are
\be
\label{eq:dnI22insp}
I_{2\pm 2}^{(q), {\rm insp}} = 2 \sqrt{\frac{2\pi}{5}} \eta M r^2 (\mp 2 i \omega)^q e^{\mp 2 i \varphi}.
\ee
Here $\omega\equiv \dot{\varphi} = (M/r^3)^{1/2}$ is the orbital frequency, $r=r_m(1-T/\tau_{\rm rr})^{1/4}$ is the orbital separation, $\varphi$ is the orbital phase, $\tau_{\rm rr}=(5/256)(M/\eta)(r_m/M)^4$, $T=t-t_m$, and $r_m$ is the orbital separation at the ``matching time'' $t_m$. For $t>t_m$ the quadrupole-moment derivatives are modeled as a sum of ringdown QNMs:
\be
\label{eq:dnI22ring}
I_{2\pm 2}^{(2+p), {\rm ring}} = \sum_{n=0}^{n_{\rm max}} (-\sigma^{\,}_{22n})^{p} A_{22n} e^{-\sigma^{\,}_{22n}T} ,
\ee
where $\sigma^{\,}_{lmn} = i\omega^{\,}_{lmn} + \tau_{lmn}^{-1}$, with QNM angular frequencies $\omega^{\,}_{lmn}$ and damping times $\tau^{\,}_{lmn}$ given in Ref.~\cite{berti-cardoso-will-PRD2006}. These QNMs depend on the final mass $M_f$ and the dimensionless spin parameter $a_f$ of the BH merger remnant and are determined by NR simulations [eg., the fits in Eqs.~(C5)-(C6) of \cite{baker-etal-PRD2008-gravlradcharacteristics}]. The coefficients $A_{22n}$ are determined by matching Eqs.~\eqref{eq:dnI22insp}-\eqref{eq:dnI22ring} at $t=t_m$ for $2 \leq (q,p+2) \leq n_{\rm max}+2$. Substituting these relations into ${h}^{\rm mem}$ yields a simple expression for the memory's evolution:
\be
\label{eq:hmemMWM}
{h}^{\rm mem}_{\rm MWM} = \frac{8\pi M}{r(T)} H(-T) + \left\{ \! \frac{8\pi M}{r_m} + \frac{1}{\eta M} \sum_{n,n'=0}^{n_{\rm max}} \frac{\sigma^{\,}_{22n}\sigma_{22n'}^{\ast}}{\sigma^{\,}_{22n} + \sigma_{22n'}^{\ast}} A^{\,}_{22n} A_{22n'}^{\ast} \left[ 1 - e^{-(\sigma^{\,}_{22n} + \sigma_{22n'}^{\ast})T} \right] \! \right\} \! H(T),
\ee
where $H(T)$ is the Heaviside function. Choosing $n_{\rm max}=2$ and $r_m=3M$ (corresponding to the light-ring of a Schwarzschild BH), yields a final saturation value of the memory $\Delta {h}^{\rm mem}_{\rm MWM} \approx 16$.
\begin{figure*}[t]
$
\begin{array}{cc}
\includegraphics[angle=0, width=0.47\textwidth]{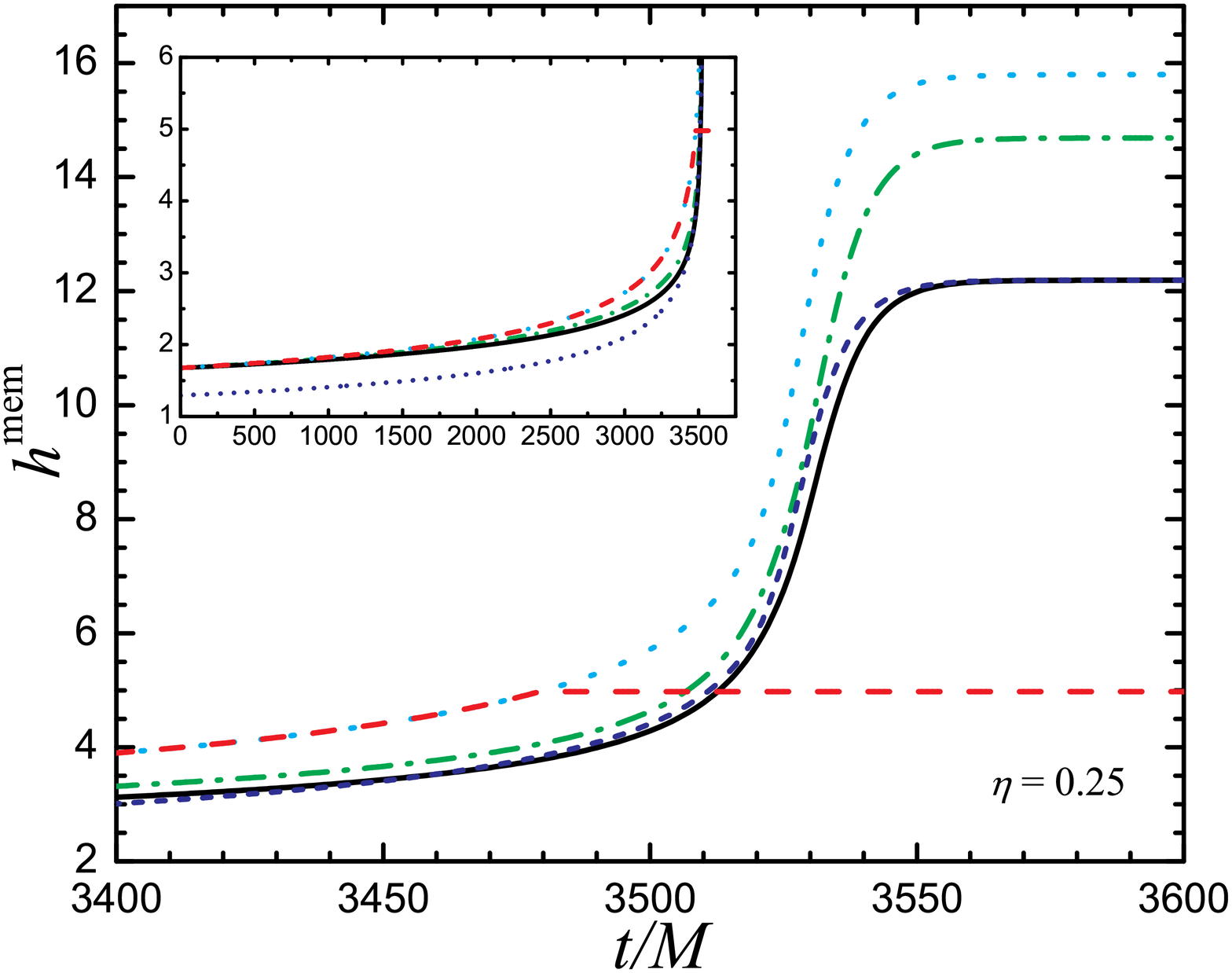} &
\includegraphics[angle=0, width=0.485\textwidth]{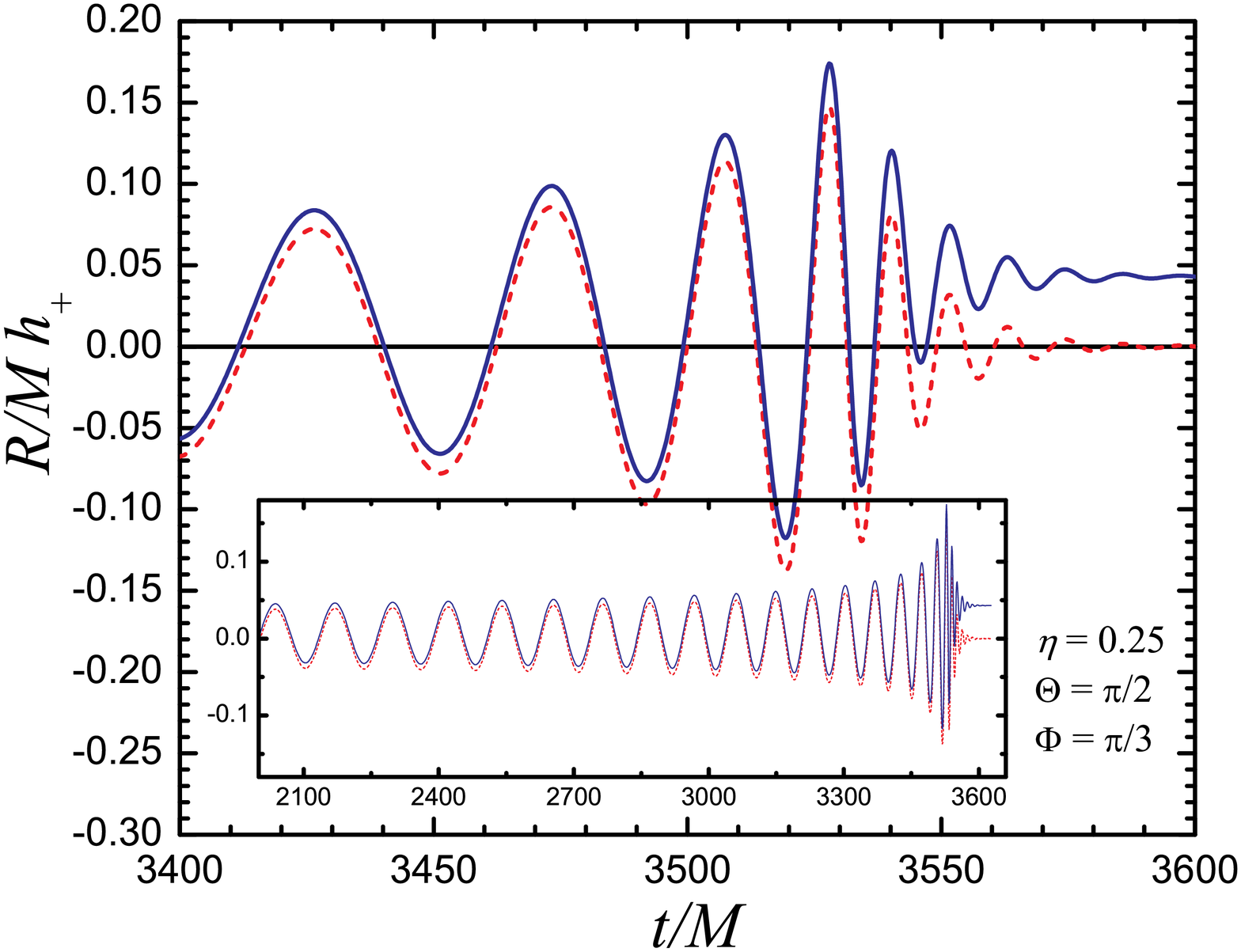}
\end{array}
$
\caption{\label{fig:hmem}The left plot shows the evolution and saturation of the memory ${h}^{\rm mem}$ near the merger time. The solid (black) line uses the full-EOB formalism calibrated to numerical relativity simulations as in Ref.~\cite{damour-nagar-jena}. The dashed-dotted (green) line uses this same formalism but without the EOB amplitude correction factors $f_{22}^{\rm NQC} F_{22}$. The dotted (cyan) curve is the minimal-waveform model. The short-dashed (blue) curve is the minimal-waveform model multiplied by a ``fudge factor'' $\approx 0.77$ so that it matches the full-EOB curve at late times. The red (dashed) curve is similar to the treatment in \cite{kennefick-memory}: the inspiral memory is truncated at an orbital separation $r=5M$.  Note that the PN/EOB corrections tend to reduce the memory's magnitude. The right plot shows the $h_{+}$ polarization with memory (sold/blue) and without (dashed/red). This is computed using the full-EOB model. The inset plots show the early-time evolution. Both plots are for an equal-mass binary with the matching to the ringdown signal at $t_m/M \approx 3522$.}
\end{figure*}

In addition to the MWM, we can model the evolution of $I_{2\pm2}$ using the EOB approach of Ref.~\cite{damour-nagar-jena}, where the freely adjustable EOB parameters are determined by fitting to simulations from the Jena and Caltech/Cornell NR groups. The main differences between the MWM and the full-EOB calculation are: (i) the EOB equations of motion are solved to determine the binary separation and orbital frequency during the inspiral, (ii) PN correction factors to the amplitude of $I^{(q), {\rm insp}}_{2\pm2}$ are included as in Ref.~\cite{damour-nagar-jena}, and (iii) we match to 5 QNMs at 5 points near the EOB-deformed light-ring for $q=p+2=2$. The results of the MWM and full-EOB model are presented in Figure \ref{fig:hmem}. Further details are presented in Ref.~\cite{favata-memory-saturation}.

These results can be used to compute the signal-to-noise ratio (SNR) for the memory signal \cite{favata-memory-saturation}. For initial LIGO the memory will be undetectable, but advanced LIGO may have a slim chance of detecting the memory from nearby BBH mergers (${\rm SNR}\approx 8$ for a $50 M_{\odot}/50 M_{\odot}$ binary at $20$ Mpc). LISA has good prospects for detecting the memory from supermassive BBH mergers at large redshifts: for example a $10^5 M_{\odot}/10^5 M_{\odot}$ binary at redshift $z=2$ will have a ${\rm SNR} \approx 9$.
\section{\label{sec:recoilmem}Signature of radiation recoil in the gravitational-wave signal}
Numerical relativity (NR) calculations of gravitational radiation recoil have found that BHs can receive kicks as large as $\sim 175 \text{ km/s}$ for nonspinning BHs \cite{gonzalez-maxnonspinningkick}, $\sim 450 \text{ km/s}$ for equal-mass BHs with spins anti-aligned along the orbital angular momentum \cite{pollney-spinalignedrecoilsPRD}, and $\sim 4000 \text{km/s}$ for the ``super-kick'' configuration (spins anti-aligned in the orbital plane) \cite{campanelli-PRL07-recoil}. These moderate to large recoils can have interesting electromagnetic signatures
\cite{favata-kickconsequences-ApJL,loeb-kicksignaturesPRL,
volonteri-madu-offnucAGN,
devecchi-xrayrecoilingBHs,
mohayaee-gammarayDMrecoiledBHs,
schnittman-krolik-IRmergerafterglow,
shields-bonning-BHflares,
haiman-etal-LISA7}, and can also leave signatures in the GW signal. This GW signature could confirm the nature of the recoil (anisotropic GW emission vs.~3-body recoil) and determine the kick magnitude and direction.

During the inspiral the center-of-mass recoil is a negligible high-PN-order effect; but if the masses and spins are accurately determined from the inspiral waves, the final kick after the merger can be inferred from NR simulations. Following the merger the recoil grows large enough to leave potentially detectable effects on the GW signal. One of these effects is an additional contribution to the GW memory. This ``kick memory'' arises from Thorne's memory formula [Eq.~\eqref{eq:hijlinmem}] for a system of $N$ unbound masses:
\be
\label{eq:kickmem}
\Delta h_{jk}^{\rm TT} = \sum_{A=1}^{N-1} \frac{4 E_A^{\rm graviton}}{R} \left[ \frac{n_A^j n_A^k}{1-{\bm n}_A \cdot {\bm N}} \right]^{\rm TT} + \frac{4 M_f}{R \sqrt{1-V_{\rm kick}^2}} \left[ \frac{V^j_{\rm kick} V^k_{\rm kick}}{1-{\bm V}_{\rm kick} \cdot {\bm N}} \right]^{\rm TT} .
\ee
Here the sum in the first term is over the $N-1$ individual gravitons radiated throughout the entire coalescence; this is the Christodoulou memory [$\delta h_{jk}^{\rm TT}$; Eq.~\eqref{eq:hTTnonlinmem}]. The second term is the memory ($\Delta h_{jk}^{\rm TT, kick}$) from the remaining particle in the system---the kicked BH with final mass $M_f$ and recoil velocity $V_j^{\rm kick}$. Since the Christodoulou memory roughly scales with the radiated energy, $\Delta h^{\rm Chris.} \sim 4 \Delta E_{\rm rad}/R$, we can see that the ratio of the two memories scales like
\be
\frac{\Delta h^{\rm kick}}{\Delta h^{\rm Chris.}} \sim \frac{V_{\rm kick}^2}{\Delta E_{\rm rad}/M} \sim 3 \times 10^{-3} \left(\frac{V^{\rm kick}}{3000 \text{ km/s}}\right)^2 \left( \frac{3\%}{\Delta E_{\rm rad}/M} \right) .
\ee
Since plausible SNRs for the Christodoulou memory are $< 100$, it seems unlikely that the ``kick memory'' will be easily detected.

Prospects are better for detecting a Doppler shift of the QNM oscillations. Measuring this Doppler shift relies on accurately measuring the masses and spins from the inspiral, and using NR simulations to determine the rest-frame QNMs. Comparison with the observed QNMs then determines the line-of-sight (LOS) recoil velocity, $V_{\rm LOS} = {\bm V}_{\rm kick} \cdot {\bm N}$. The accuracy with which $V_{\rm LOS}$ can be determined then depends on how accurately the QNMs can be measured. Using Eq.~(7.2) of Ref.~\cite{berti-cardoso-will-PRD2006}, we can roughly estimate the relative error in the LOS recoil: $\sigma_{V_{\rm LOS}}/V_{\rm LOS} \sim 2/(\rho V_{\rm LOS})$, where $\rho$ is the SNR from the QNM signal. For large SNRs and moderately large kicks, this implies a relative error of a few tens of percent.
\section{\label{sec:conclusions}Conclusions}
While previous works have computed the memory only during the inspiral, here the memory is computed through the merger and ringdown. The calculation has used a simple, fully-analytic model for the coalescence as well as a previously developed effective-one-body (EOB) approach calibrated to numerical simulations. Prospects for detecting the memory are poor for initial and advanced LIGO, but are promising for supermassive BH mergers that LISA will detect with high signal-to-noise ratios ($\gtrsim 1000$). Binaries that recoil after merger also show a linear memory, as well as a Doppler shift of the quasi-normal-mode frequencies. These effects will be difficult to detect except in strong sources with moderately large kicks.

Although challenging, it will be important for numerical relativity simulations to test the semi-analytic computations of the memory discussed here. Future Mock LISA Data Challenges \cite{mldc-cqg07} should also consider including waveforms with memory in their data sets. This will provide a better understanding of LISA's ability to measure this interesting effect.
\ack
This research was supported in part by the National Science Foundation under
Grant No. PHY05-51164. I am grateful to several participants of the $7^{\rm th}$ LISA Symposium for helpful discussions concerning this work.
\section*{References}
\bibliographystyle{iopart-num}
\bibliography{lisa7confprocV2}
\end{document}